\documentclass[aps,pra,showpacs,superscriptaddress,twocolumn,eqsecnum]{revtex4}
\usepackage{graphicx}
\usepackage{bm,dsfont}
\usepackage{amsmath,amssymb,amsthm}
\begin{document}
\preprint{}
\title{SU(2)-invariant depolarization of quantum states of light}
\author{\'{A}ngel Rivas}
\affiliation{Departamento de F\'isica Te\'orica I, Facultad de Ciencias F\'isicas, Universidad Complutense, 28040 Madrid, Spain}
\author{Alfredo Luis}
\email{alluis@fis.ucm.es}
\homepage{http://www.ucm.es/info/gioq}
\affiliation{Departamento de \'{O}ptica, Facultad de Ciencias
F\'{\i}sicas, Universidad Complutense, 28040 Madrid, Spain }
\date{\today}

\begin{abstract}
We develop a SU(2)-invariant approach to the depolarization of quantum systems as the effect of random unitary SU(2) transformations. From it we derive a SU(2)-invariant Markovian master equation. This is applied to several quantum states examining whether nonclassical states are more sensible to depolarization than the classical ones. Furthermore, we show that this depolarization model provides a non-trivial generalization of depolarization channels to states of arbitrary dimension.
\end{abstract}

\pacs{03.65.Yz, 42.50.Ar, 42.25.Ja, 42.50.Lc}


\maketitle

\section{Introduction}

Polarization is a key ingredient of light. Besides being a fundamental manifestation of coherence, this is of practical relevance in areas such as optical communications, interferometry, and metrology, both in classical and quantum domains \cite{Pol,Pol2,Agarwal,Scully,Nielsen}. In all of these applications the effect of depolarization mechanisms is naturally of importance. For example it is worth examining how sensitive to depolarization are quantum states with metrological interest.

In quantum optics the basic observables which characterize polarization are the Stokes operators, whose mean values correspond to the classical Stokes parameters \cite{Pol2,Agarwal}. The three Stokes operators are formally equivalent to an angular momentum as they fulfill the $\mathfrak{su}(2)$ commutation algebra. Thus, the polarization properties of some state of light can be visualized as a probability on a sphere, the Poincar\'e sphere \cite{Pol,Pol2,Agarwal}, on a three-dimensional space generated by the three Stokes operators. Hence, field states differing by a SU(2) transformation (i.e. a rotation on the Poincar\'e sphere) are statistically equivalent. This is to say that polarization properties must be independent of the polarization basis. From this perspective we shall consider depolarization processes that attain the same form in all polarization bases. Therefore, the analysis is based just on SU(2) symmetry without any further specific assumption.

In this regard, all real optical systems are unavoidably non-deterministic in polarization to a larger or smaller extent because of microscopic inhomogeneities and anisotropies. These are specially relevant in the optical domain because of the smallness of the wavelength. Thus, real optical devices must be more properly represented by an ensemble of random transformations. Here we are interested in the structure, symmetries, and invariance properties of the depolarization caused by randomness.

Depolarization can be understood as the transfer of energy from polarized states to the unpolarized state \cite{Pol2}. We focus on depolarization caused exclusively by random, linear, unitary, energy conserving processes, which are represented by random SU(2) transformations \cite{QBS}. Attenuation does not provide any further insight since isotropic losses do not alter polarization, while anisotropic losses (diattenuators) are actually polarizing devices.

In this work the following points are discussed in detail.

\noindent i/  We formulate a model for SU(2)-invariant depolarization processes in two forms: first via a finite form and second by deriving a Markovian master equation.

\noindent ii/ This provides us with a framework to study the robustness of polarization independently of the basis, in such a way that all states connected by a SU(2) transformation are treated on equal footing.

\noindent iii/ We show that our model of depolarization cannot be written in general with the typical form of a depolarization channel commonly used in quantum information theory \cite{Nielsen}. That is only possible for one-photon states (qubits).

\noindent iv/ Remarkably, for the usual degree of polarization, the rate of depolarization is the same independently of the field state. However, by employing more sophisticated degrees of polarization we find that polarization of relevant SU(2) nonclassical states (such as twin photon number or NOON states) is more fragile than for the case of SU(2) coherent states.

\noindent v/ Notably, this rate of decay for polarization is approximately independent of the number of photons.

The paper is organized as follows. In Sec. II we recall the main tools required to deal with quantum polarization. Then in Sec. III we model depolarization as the result of random SU(2) transformations and we study their effect on some degrees of polarization. After considering its infinitesimal form, we derive in Sec. IV a SU(2) invariant master equation examining its most interesting features. We apply this to some examples in Sec. V, examining whether quantum states are more sensible to depolarization than classical states. As mentioned, among other consequences we show that SU(2) invariance provides a suitable generalization of depolarization channels to arbitrary photon numbers
\cite{dch}.

Most of the  results reported here apply equally well to quantum and classical optics provided we replace the quantum density matrix $\rho$ by the classical cross-spectral density tensor $C_{ij} = \langle E (\mathbf{r}_i, \omega) E^\ast (\mathbf{r}_j, \omega) \rangle$ in the space-frequency domain for example.

\section{Quantum polarization}

Next we recall the main facts about quantum polarization
required below.

\subsection{Stokes operators and SU(2) transformations}

Quantum polarization is conveniently described in terms of
the Stokes operators
\begin{equation}
\label{So}
\begin{array}{ll}
S_0 := a^\dagger_1 a_1 + a^\dagger_2 a_2, &
S_x := a^\dagger_2 a_1 + a^\dagger_1 a_2, \\
S_y := {\rm i} \left ( a^\dagger_2 a_1 - a^\dagger_1 a_2 \right ), &
S_z := a^\dagger_1 a_1 - a^\dagger_2 a_2 ,
\end{array}
\end{equation}
where $a_{1,2}$ are the complex amplitudes of two field modes. For $\bm{S} = (S_x, S_y, S_z )$, it holds that
\begin{equation}
\label{S2}
\bm{S}^2 = S_0 \left ( S_0 + 2 \right ).
\end{equation}
The Stokes operators fulfill the commutation relations of an
angular momentum \cite{SCH}
\begin{equation}
\label{cr}
[ S_k ,S_\ell ] = 2 {\rm i} \sum_{m=x,y,z} \epsilon_{k\ell m}
S_m , \qquad [S_0, \bm{S} ] = \bm{0} ,
\end{equation}
where $\epsilon_{k\ell m}$ is the fully antisymmetric tensor with $\epsilon_{xyz} =1$.

Because the commutation with the total number of photons $S_0$,
the action of the Stokes operators leaves invariant the subspaces
$\mathcal{H}_n$ with fixed total photon number $n$ (and dimension $n+1$). These subspaces
are spanned by the photon-number states $|m, n-m \rangle$ with $m$
photons in mode $a_1$ and $n-m$ photons in mode $a_2$. The whole
Hilbert space can decompose as a direct sum of these subspaces
$\mathcal{H} = \bigoplus_{n=0}^\infty  \mathcal{H}_n$.

The Stokes operators are also the infinitesimal generators of SU(2) unitary transformations $U$
\begin{equation}
U(\bm{u}) = \exp \left ( {\rm i} \bm{u} \cdot \bm{S} \right ),
\end{equation}
where $\bm{u}$ is a three-dimensional real vector, which produces
a rotation $R$ of $\bm{S}$ \cite{ACGT}
\begin{equation}
\label{rot}
U^\dagger (\bm{u}) \bm{S} U (\bm{u}) = R (\bm{u}) \bm{S} ,
\end{equation}
with $R^{\rm t} R = R R^{\rm t} = \mathds{1}$, where the superscript ${\rm t}$ denotes matrix
transposition. The vector $\bm{u}$ expresses both the axis
and angle $\alpha$ of rotation, with $\alpha = 2 | \bm{u}|$.
In practical terms, SU(2) transformations describe basic
and ubiquitous optical devices, which are all the energy-conserving linear processes such a lossless beam splitters,
phase shifters, phase plates, and basic interferometers.

\subsection{SU(2) coherent states and polarization distribution}

Complete information about polarization properties is given
by a polarization probability distribution on the Poincar\'{e}
sphere. Maybe the best behaved expressions are provided by the
SU(2) $Q$ function defined as \cite{ACGT,LU1}
\begin{equation}
\label{SUQ}
Q(\bm{\Omega}) := \sum_{n=0}^\infty \frac{n+1}{4 \pi}
\langle n, \bm{\Omega} | \rho | n, \bm{\Omega} \rangle,
\end{equation}
where $\rho$ is the density matrix for the two-mode field,
and $| n, \bm{\Omega} \rangle$ are the SU(2) coherent states,
expressed in the photon-number basis as
\begin{eqnarray}
| n, \bm{\Omega} \rangle & := & \sum_{m=0}^n
\left(
\begin{array}{c}
n \\ m
\end{array}\right )^{1/2}
 \left ( \sin
\frac{\theta}{2} \right )^{n-m} \left ( \cos
\frac{\theta}{2} \right )^m
\nonumber  \\
& & \times {\rm e}^{-{\rm i} m \phi} |m , n-m \rangle ,
\end{eqnarray}
so that $\theta$ and $\phi$ represent the polar and the
azimuthal angles, respectively, of the Poincar\'{e} sphere.
The SU(2) coherent states can be defined by the following
eigenvalue equation
\begin{equation}
\label{cseq}
\bm{\Omega} \cdot \bm{S} | n, \bm{\Omega} \rangle = n
| n, \bm{\Omega} \rangle , \quad
\bm{\Omega} := \begin{pmatrix} \sin \theta \cos \phi \\ \sin
\theta \sin \phi \\ \cos \theta \end{pmatrix} .
\end{equation}

It is worth noticing in Eq. (\ref{SUQ}) that the matrix
elements of $\rho$ connecting subspaces $\mathcal{H}_n$ of
different total photon number $n$ do not contribute to
$Q(\bm{\Omega})$. This is consistent with the fact that
polarization and intensity are in principle independent
concepts: the form of the ellipse described by the electric
vector (polarization) versus the size of the ellipse
(intensity). This is also consistent with the commutation
of any function of the Stokes operators $f(\bm{S})$ with
the total number of photons, $[f(\bm{S}), S_0]=0$, so that
the matrix elements of $\rho$ connecting subspaces of
different total photon number $n$ do not contribute to
$\langle f(\bm{S}) \rangle$.

\subsection{Polarization fluctuations and degree of polarization}

Most analyses of polarization focus exclusively on the Stokes
parameters, which are the mean values of the Stokes operators
$s_0 = \langle S_0 \rangle$ and $\bm{s} = \langle \bm{S}
\rangle$. Thus, the standard (first order)
degree of polarization is defined as
\begin{equation}
\label{dp}
P_s := \frac{|\bm{s}|}{s_0} .
\end{equation}
This is essentially a SU(2)-invariant assessment of polarization
fluctuations via the sum of the variances of any three orthogonal
Stokes components $S_{1,2,3}$  \cite{DE},
\begin{equation}
\label{svSo}
( \Delta \bm{S} )^2 := ( \Delta S_1 )^2 + ( \Delta S_2 )^2 +
( \Delta S_3 )^2 ,
\end{equation}
which can be expressed as
\begin{equation}
\label{svSo2}
( \Delta \bm{S} )^2 = \langle S_0 \left ( S_0 + 2 \right )
\rangle - P_s^2 \langle S_0 \rangle^2 .
\end{equation}
This implies that polarization uncertainty is bounded both from
above and below
\begin{equation}
\label{aab}
\langle S_0 \left ( S_0 + 2 \right ) \rangle \geq ( \Delta
\bm{S} )^2 \geq 2 \langle S_0  \rangle ,
\end{equation}
where the minimum $\Delta \bm{S}$ is reached by SU(2) coherent
states since they have the larger $P_s$ possible.

The degree of polarization (\ref{dp}) is not always fully
satisfactory since $P_s$ is defined solely in terms of the
first moment of the Stokes operators. Thus it cannot reflect the
basic quantum polarization properties defined in terms of
higher order moments, such as polarization squeezing
\cite{LK,BM}. A more complete degree of polarization $P_Q$
can be defined in terms of the distance $D$ between the
SU(2) $Q$ function and the uniform SU(2) $Q$ function
$1/(4 \pi)$ describing fully unpolarized light \cite{LU1} as
\begin{equation}
\label{nddp}
P_Q := \frac{D}{1+D} = 1 - \frac{1}{4 \pi}
\Sigma,
\end{equation}
where
\begin{equation}
\label{dS}
\Sigma := \frac{1}{\int d \Omega  \left [ Q(\bm{\Omega} )
\right ]^2 },
\end{equation}
with
\begin{equation}
\label{dD}
D:= 4 \pi \int d \Omega \left [ Q(\bm{\Omega} ) - \frac{1}{4 \pi}
\right ]^2 =  4 \pi \int d \Omega  Q^2(\bm{\Omega}) - 1,
\end{equation}
and $d \Omega = \sin \theta d \theta d \phi$. The function
$\Sigma$ can be interpreted as an effective area of the
Poincar\'{e} sphere where the $Q$ function is different from
zero. Since each point of the Poincar\'{e} sphere
represents a different polarization state, $\Sigma$ assesses
how many polarization states have nonvanishing probability
to appear in a given field state. The contribution to
$\Sigma$ of each point being properly weighted by its
probability.

Both Eqs. (\ref{dp}) and (\ref{nddp}) are invariant under
deterministic SU(2) transformations so that the states
$\rho$ and $U (\bm{u} ) \rho U^\dagger (\bm{u} )$ have
the same $P_{s}$ and $P_Q$.

\section{Depolarization in finite form}

As aforementioned, real optical systems are unavoidably non-deterministic
to a larger or smaller extent because of practical imperfections,
randomness, inhomogeneities, and so on. Concerning purely depolarization processes we can focus on linear
energy-conserving devices that can be represented by unitary SU(2)
transformations $U (\bm{u})$ that may occur at random following a
given time-dependent probability distribution $p(\bm{u},t)$. Irrespective
of whether the effect is larger or smaller, here we are
interested mainly in the structure, symmetries, and invariance of
the depolarization process, and the main consequences that can be
derived from such basic traits.

The transformed state $\rho (t)$  after a time $t$ experiencing
random unitary transformations can be related with the original
state $\rho (0)$ at $t=0$ as
\begin{equation}
\label{fd}
\rho (t) = \mathcal{E}_{(t,0)} [\rho (0)] = \int d^3 \bm{u} p (\bm{u},t)
U (\bm{u}) \rho (0) U^\dagger (\bm{u}) ,
\end{equation}
where $p (\bm{u},t)$ is the probability that the SU(2)
transformation $U (\bm{u})$ occurs at time $t$,
\begin{equation}
p (\bm{u},t) \geq 0, \qquad \int d^3 \bm{u} p (\bm{u},t) =1 .
\end{equation}
Moreover, we take $p (\bm{u},0)=\delta^3(\bm{u})$ or another similar condition such that the equality in Eq. \eqref{fd} is also satisfied at $t=0$.

Equation (\ref{fd}) is a unital transformation \cite{UN}, so that the identity $\mathds{1}_n$ in each subspace
$\mathcal{H}_n$ is a fixed point, i.e. $\rho (0) =\mathds{1}_n/(n+1)
\rightarrow \rho (t) = \mathds{1}_n/(n+1)$ for every $p (\bm{u},t)$.
In the polarization context this is depolarization
with zero polarizance \cite{Pol2}. The unital character implies that the quantum-state purity
cannot increase, $\textrm{Tr} [ \rho^2 (t ) ] \leq
\textrm{Tr}[ \rho^2 (0 )]$. This can be seen by applying the
Cauchy-Schwarz inequality $| \textrm{Tr} (A B^\dagger) |
\leq \sqrt{\textrm{Tr} (A A^\dagger) \textrm{Tr} (B B^\dagger)}$
to $\textrm{Tr}[ \rho^2 (t )]$,
\begin{align}
\textrm{Tr} [ \rho^2 (t)] =& \int d^3 \bm{u} d^3 \bm{u}^\prime
p (\bm{u} ,t) p (\bm{u}^\prime ,t)  \\
&\times\textrm{Tr} \left [ U
(\bm{u}) \rho (0) U^\dagger (\bm{u}) U (\bm{u}^\prime )
\rho (0) U^\dagger (\bm{u}^\prime) \right ],\nonumber
\end{align}
with $A = U (\bm{u}) \rho (0) U^\dagger (\bm{u})$ and $B =
U (\bm{u}^\prime ) \rho (0) U^\dagger (\bm{u}^\prime )$. For
deterministic transformations $p (\bm{u},t) = \delta^3 [
\bm{u}- \bm{u}_0 (t)]$ it holds $\textrm{Tr} [ \rho^2 (t )] =
\textrm{Tr} [ \rho^2 (0 )]$, which is a necessary condition
for monotonic behavior of the degree of polarization $P_s$
under random transformations \cite{RE}.

\subsection{Decrease of degrees of polarization}

Let us show explicitly that the degrees of polarization
(\ref{dp}) and (\ref{nddp}) can never increase under the effect
of random SU(2) transformations. Concerning $P_s$  we have that
\begin{align}
\bm{s} (t) &= \int d^3 \bm{u} p(\bm{u},t) \langle U^\dagger
(\bm{u}) \bm{S} U (\bm{u}) \rangle \nonumber \\
&= \int d^3 \bm{u}
p(\bm{u},t) R(\bm{u}) \bm{s}(0) ,
\end{align}
where $R(\bm{u})$ is the rotation associated with $U (\bm{u})$
in Eq. (\ref{rot}), while
\begin{equation}
s_0 (t) = \int d^3 \bm{u} p(\bm{u},t) \langle U^\dagger
(\bm{u}) S_0 U (\bm{u}) \rangle = s_0 (0) ,
\end{equation}
because $[S_0 , U (\bm{u})]=0$, $U^\dagger (\bm{u}) U (\bm{u}) = \mathds{1}$,
and $\int d^3 \bm{u} p(\bm{u},t) =1$. Thus $P_s (t) \leq P_s (0)$
since
\begin{align}
P_s (t) &= \frac{|\bm{s} (t) |}{s_0 (t)} = \frac{1}{s_0 (0)}
\left | \int d^3 \bm{u} p(\bm{u},t) R(\bm{u}) \bm{s}(0) \right | \nonumber\\
&\leq
\frac{1}{s_0 (0)}
\int d^3 \bm{u} p(\bm{u},t) \left | R(\bm{u}) \bm{s}(0) \right |
= P_s (0),
\end{align}
where we have used that  $\left | R(\bm{u}) \bm{s}(0) \right |
= \left | \bm{s}(0) \right | = P_s (0) s_0 (0)$.

This further implies that the total uncertainty of the Stokes
operators cannot decrease after a random SU(2) transformation. This can be seen
by using Eq. (\ref{svSo2}) and taking into account that $\langle
S_0 \left ( S_0 + 2 \right ) \rangle$, $\langle S_0 \rangle$ are
invariant under SU(2) transformations (deterministic and random),
so that $P_s (t) \leq P_s (0)$ implies that $( \Delta \bm{S} )^2(t)\geq( \Delta \bm{S} )^2(0)$.

Concerning $P_Q$, we adapt to this context a similar result derived
in classical optics \cite{RL}. To this end we note that
from Eqs. (\ref{rot}) and (\ref{cseq})
\begin{equation}
U^\dagger (\bm{u}) | n, \bm{\Omega} \rangle = | n, R^{\rm t} (\bm{u} )
\bm{\Omega} \rangle,
\end{equation}
so that from Eq. (\ref{fd})
\begin{equation}
Q(\bm{\Omega},t) = \int d^3 \bm{u} p (\bm{u},t)
Q \left [ R^{\rm t} (\bm{u}) \bm{\Omega},0 \right ] .
\end{equation}
For evaluating $P_Q (t) $ we need
\begin{equation}
\int d \Omega Q^2 (\bm{\Omega},t) = \int d \Omega \left \{
\int d^3 \bm{u} p (\bm{u},t) Q \left [ R^{\rm t} (\bm{u}) \bm{\Omega},0
\right ] \right \}^2 .
\end{equation}
Since the square is a convex function,
\begin{align}
\int d \Omega & \left \{ \int d^3 \bm{u} p (\bm{u},t)
Q \left [ R^{\rm t} (\bm{u}) \bm{\Omega},0 \right ] \right \}^2 \nonumber \\
&\leq
\int d^3 \bm{u} p (\bm{u},t) \int d \Omega Q^2 \left [ R^{\rm t}
(\bm{u}) \bm{\Omega},0 \right ].
\end{align}
Taking into account that $Q [ R^{\rm t} (\bm{u}) \bm{\Omega},0 ]$ is
the polarization distribution of the state $U (\bm{u}) \rho (0)
U^\dagger (\bm{u})$ and that $\int d \Omega Q^2 ( \bm{\Omega},0)$
is invariant under SU(2) transformations we get $\int d \Omega
Q^2 [ R^{\rm t} (\bm{u}) \bm{\Omega},0 ] = \int d \Omega Q^2 (
\bm{\Omega},0 )$ that does not depend on $\bm{u}$ and then
\begin{equation}
\int d^3 \bm{u} p (\bm{u},t) \int d \Omega
Q^2 \left [ R^{\rm t} (\bm{u}) \bm{\Omega},0 \right ]=
\int d \Omega Q^2 (  \bm{\Omega},0) ,
\end{equation}
therefore
\begin{equation}
\int d \Omega Q^2 (  \bm{\Omega},t) \leq \int d \Omega Q^2 (
\bm{\Omega},0) ,
\end{equation}
which in turn implies that $\Sigma (t) \geq \Sigma (0)$ and
finally $P_Q (t) \leq P_Q (0)$.

\subsection{SU(2) invariance}

As we are interested in SU(2)-invariant depolarization next
we study the properties that $p (\bm{u},t)$ must satisfy to fulfill this requirement. The idea of SU(2) invariance is that
there are no privileged polarization states and all
points of the Poincar\'{e} sphere are on an equal footing. This
is to say that if depolarization transforms $\rho (0)$ into
$\rho (t)$, then it should transform $V \rho (0) V^\dagger$ into
$V \rho (t) V^\dagger$, where $V$ is any SU(2) transformation.
This means that Eq. (\ref{fd}) and
\begin{equation}
\rho (t) = \int d^3 \bm{u} p (\bm{u},t) V^\dagger U(\bm{u})V
\rho (0) V^\dagger U^\dagger (\bm{u}) V ,
\end{equation}
should hold simultaneously. Since $V^\dagger U(\bm{u})V= U
(R^{\rm t} \bm{u})$, where $R$ is the rotation associated with $V$
in Eq. (\ref{rot}), we get that the condition for SU(2) invariance
is
\begin{align}
\label{ui}
\int d^3 \bm{u} &p (R \bm{u},t) U (\bm{u}) \rho (0)
U^\dagger (\bm{u}) \nonumber \\
&= \int d^3 \bm{u} p (\bm{u},t) U (\bm{u})
\rho (0) U^\dagger (\bm{u}) ,
\end{align}
which holds provided that $p (R \bm{u},t) = p (\bm{u},t)$ for
arbitrary $R$. Therefore, $p(\bm{u},t)$ must depend just on
the modulus of $\bm{u}$, i.e. $p (\bm{u},t) = p ( u = |\bm{u}|,t)$.
This is to say that there is complete isotropy for the axes of
the rotations $R(\bm{u})$ all axes being equally probable.

\subsection{Invariant state}

The invariant state $\rho (t) = \rho (0)$ of transformation
(\ref{fd}) under SU(2) invariance is given by the solution of
\begin{equation}
\label{fpe}
\rho_{\rm I} = \int_0^{\pi} du u^2 p(u,t) \int d \Omega U(u,
\bm{\Omega}) \rho_{\rm I} U^\dagger (u,\bm{\Omega}) ,
\end{equation}
where we have used spherical coordinates $\bm{u} = u \bm{\Omega} $, $d^3
\bm{u} = u^2 d u d \Omega$ (recall that $\alpha=2u$ is the angle of rotation,
so $u$ runs from 0 to $\pi$). As we have noted above, the
transformation is unital and all the identities $\mathds{1}_n$ in
$\mathcal{H}_n$ are solutions. More specifically we have
\begin{equation}
\label{ss}
\rho_{\rm I} = \bigoplus_{n=0}^\infty p_n \frac{1}{n+1} \mathds{1}_n,
\end{equation}
where $p_n$ are arbitrary parameters satisfying
\begin{equation}
\sum_{n=0}^\infty p_n = 1 , \quad p_n \geq 0.
\end{equation}
This is fully unpolarized light with uniform polarization
distribution $Q_{\rm I} (\bm{\Omega}) = 1/(4 \pi)$.

Moreover, light states with uniform polarization
distribution  are the only fixed points concerning
polarization effects as it can be seen from very simple
geometrical arguments. This is because Eq. (\ref{fpe})
implies that $Q (\bm{\Omega})$ must remain invariant
under the application of rotations with arbitrary axis [we
are assuming $p(u,t) \neq 0$ for at least one $u \neq \pi N$
for integer $N$]. This is only possible if the distribution
on the sphere is uniform $Q_{\rm I} (\bm{\Omega}) = 1/(4 \pi)$.
Incidentally, this analysis recalls the definition of
unpolarized light via transformation properties \cite{np}.

\section{Master equation}

Here, we derive a Markovian master equation for a SU(2)-invariant depolarization process as described in the previous section. The derivation of this master equation relies on several assumptions.
\begin{enumerate}
\item We consider the finite transformation Eq. (\ref{fd}) and assume it fulfils the Markovian (semigroup) property, $\mathcal{E}_{\tau_2+\tau_1}=\mathcal{E}_{\tau_2}\mathcal{E}_{\tau_1}$. Here $\tau$ denotes the difference between the final and initial time, with $\mathcal{E}_{\tau_1}:=\mathcal{E}_{(t_{1},0)}$ and $\mathcal{E}_{\tau_2}:=\mathcal{E}_{(t_{2},t_1)}$.
\item The evolution is continuous on $t$, which is sufficient to be differentiable if the above semigroup condition is fulfilled \cite{Libro}.
\item The depolarization process is SU(2)-invariant, so $p (\bm{u},t) = p(u,t)$, and the initial condition reads $p(u,0)=\delta(u)/(4\pi u^2)$.
\end{enumerate}

Under the first and second assumptions, the time evolution can be written as a master equation,
\begin{equation}
\frac{\rho(t+h)-\rho(t)}{h}=\frac{\mathcal{E}_{h}-\mathds{1}}{h}\rho(t)\xrightarrow{h\rightarrow0}\frac{d\rho(t)}{dt}=\mathcal{L}\rho(t), \end{equation}
where the generator reads
\begin{equation}
\label{generator}
\mathcal{L}:=\lim_{h\rightarrow0}{\frac{\mathcal{E}_{h}-\mathds{1}}{h}}.
\end{equation}

The third assumption allows us to obtain a more concrete form for the generator. We may expand
\begin{equation}
\label{Eh}
\mathcal{E}_{h}(\rho)=\int_0^{\pi} du u^2 p(u,h) \int d \Omega U(u,
\bm{\Omega}) \rho U^\dagger (u,\bm{\Omega}),
\end{equation}
at first order on $h$. Thus, on one hand we have
\begin{equation}
\label{pe}
p(u,t)\simeq p(u,0)+p'(u,0)t,
\end{equation}
where $p'(u,0):=\left.\frac{\partial p(u,t)}{\partial t}\right|_{t=0}$. On the other hand, by continuity, at the limit $h\rightarrow0$, the only transformations with nonvanishing probability $p(u,t\ll1) \neq 0$ are the ones
with $u \ll 1$. This allows a power-series expansion of
$U(\bm{u})$ in $\bm{u}$:
\begin{equation}
\label{se}
U(\bm{u}) \simeq  \mathds{1} + {\rm i} \bm{u} \cdot \bm{S} - \frac{1}{2} \left (
\bm{u} \cdot \bm{S} \right )^2 .
\end{equation}

Next we insert expansions \eqref{pe} and \eqref{se} in Eq. \eqref{Eh} evaluating
then the averages of $u_j$ and $u_j u_k$ according to the
distribution $p(u, h)$ at first order. For the term with the zeroth order $p(u,0)$ it is straightforward to see that we obtain the identity. For the linear term in $h$, we have that
\begin{equation}
\int d^3 \bm{u} p'(u,0) \bm{u} = \frac{\partial}{\partial t}\left[\int d^3 \bm{u} p(u,t) \bm{u}\right]_{t=0} = \bm{0},
\end{equation}
because $\int d^3 \bm{u} p(u,t) \bm{u}=\bm{0}$ as can been be performing the integration in spherical coordinates for $\bm{u}$. Similarly
\begin{equation}
\int d^3 \bm{u} p'(u,0) u_j u_k = 2 \nu \delta_{jk}
\end{equation}
where the time-independent parameter $\nu$ is defined so that
\begin{equation}
\nu := \frac{\pi}{3} \int_0^\pi du p'(u,0) u^4.
\end{equation}
Thus, by extracting the derivative out of the integral sign, it is easy to see that $\nu$ is positive since $t\geq0$ by presupposition and $\int_0^\pi du p(u,0) u^4=0$ after assumption 3 above. Then, by using the definition of the generator \eqref{generator} and keeping just the
first non-trivial order in $u$ we obtain the following master equation
\begin{equation}
\label{me}
\frac{d \rho}{dt} = \nu \left (  \sum_{j=x,y,z} 2S_j
\rho S_j - \bm{S}^2 \rho -  \rho \bm{S}^2 \right ),
\end{equation}
which has the standard form for a Markovian master equation \cite{Libro,GZ,TME} and can be equivalently written as
\begin{equation}
\label{me2}
\frac{d \rho}{dt}  = - \nu \sum_{j=x,y,z} [ S_j ,[S_j ,
\rho ]].
\end{equation}
From Eq. (\ref{me2}) we can derive an equation for the
evolution of mean values of arbitrary operators $A$,
\begin{equation}
\label{eeA}
\frac{d \langle A \rangle }{dt}  = - \nu \sum_{j=x,y,z}
\left \langle [ S_j ,[S_j , A ]] \right \rangle.
\end{equation}
This kind of master equation also arises in the analysis of the evolution of an observed
system under a nonreferring (no account is taken of the
measurement results) continuous simultaneous measurement of
the three Stokes operators $\bm{S}$ \cite{cm}.

\subsection{General solutions}

The general solution of the master equation (\ref{me2}) can be written as
\begin{equation}
\rho (t) = \exp \left ( -2\nu t \bm{S}^2 \right )
\sum_{n=0}^\infty \sum_{j=x,y,z} \frac{(2 \nu t)^n}{n!}
S_j^n \rho (0) S_j^n.
\end{equation}
Alternatively, it may be formally solved in terms of the spherical
tensor operators, or multipole operators \cite{sto}, $T_{k,q}$ that
for a spin $\bm{j}$ read
\begin{equation}
T_{k,q} = \sum_{m,m^\prime=-j}^j (-1)^{j+m} \sqrt{2k+1} \begin{pmatrix} j & k & j \cr
-m & q  & m^\prime \end{pmatrix} | j,m\rangle \langle j, m^\prime |,
\end{equation}
where $|j, m\rangle$ are the eigenstates of $\bm{j}^2$ and $j_z$, $\begin{pmatrix} j & k & j
\cr -m & q  & m^\prime \end{pmatrix} $ is the Wigner $3j$ symbol, $k$ takes the values
$0, 1, \ldots, 2j$, and $q=-k,-k+1,\ldots, k$. They are orthogonal in the sense that $\mathrm{Tr}
\left ( T_{k,q}^\dagger T_{k^\prime,q^\prime} \right ) = \delta_{kk^\prime} \delta_{qq^\prime}$
and allow an expansion of any $j$-spin density operator as,
\begin{equation}
\label{ex}
\rho = \sum_{k=0}^{2j} \sum_{q=-k}^k c_{kq} T_{k,q},
\qquad c_{kq} = \mathrm{Tr} \left ( T_{k,q}^\dagger \rho \right ).
\end{equation}
The simpler example is the case $k=1$ for which we have
\begin{equation}
T_{1,0} \propto j_z , \quad T_{1,\pm 1} \propto j_\pm = j_x \pm {\rm i} j_y .
\end{equation}
The key point for our purposes is that they satisfy the following commutation
relations
\begin{align}
[j_z, T_{k,q}] &= \hbar q T_{k,q} , \nonumber \\
[j_\pm, T_{k,q}] &= \hbar \sqrt{(k \mp q)(k \pm q+1)} T_{k,q \pm 1} ,
\end{align}
so that
\begin{equation}
\label{ty}
 \sum_{\ell =x,y,z} [ j_\ell ,[j_\ell , T_{k,q} ]] = \hbar^2 k(k+1) T_{k,q} .
\end{equation}
Taking into account that the Stokes operator behave as angular momentum operators with
$\hbar \rightarrow 2$, the spherical tensor operators  can be used to solve the master equation
(\ref{me2}) within each subspace of  total photon number $n$, which corresponds to a spin $j=n/2$.
After Eqs. (\ref{me2}), (\ref{ex}), and (\ref{ty}) we get
\begin{equation}
\label{sevo}
\frac{d c_{kq}}{dt} = - 4 \nu k(k+1) c_{kq}, \quad c_{kq} (t) =
c_{kq} (0) e^{-4 k(k+1) \nu t} .
\end{equation}
We can appreciate that the slowest decaying factor $\exp (- 8 \nu t)$
corresponding to $k=1$ is common for all photon numbers $n$ provided that
 $\langle \bm{S}\rangle \neq \bm{0}$. Therefore, in the long term all states with
 $\langle \bm{S}\rangle \neq \bm{0}$ decay at the same speed irrespective of
 the number of photons $n$. On the other hand, the states with $\langle \bm{S}
 \rangle = \bm{0}$ lack the $k=1$ term ($c_{1q}=0$) and therefore decay faster. Among them
 we can find  mixed classical states such as phase-averaged equatorial SU(2)
coherent states or the incoherent superposition of antipodal SU(2) coherent
states \cite{AL07}, as well as well-known nonclassical states such as twin-number
and NOON states.

\subsection{SU(2) invariance}

Although we have examined the SU(2) invariance in the
finite form (\ref{fd}) let us address this issue directly on
the master equation (\ref{me}). We can check that this equation is SU(2) invariant in the sense that if $\rho (t)$ is
a solution then $\tilde{\rho} (t) = V \rho (t) V^\dagger$ is
also a solution, where $V$ is an arbitrary SU(2) transformation.
This means that $\tilde{\rho}$ should satisfy the master equation
\begin{equation}
\label{ime}
\frac{d \tilde{\rho}}{dt}  = - \nu \sum_{j=x,y,z} [ S_j ,
[S_j , \tilde{\rho} ]] ,
\end{equation}
which is equivalent to
\begin{equation}
\label{iime}
\frac{d \rho}{dt}  = - \nu \sum_{j=x,y,z} [ \tilde{S}_j ,
[\tilde{S}_j , \rho ]] ,
\end{equation}
where $\tilde{\bm{S}} = R \bm{S} = V^\dagger \bm{S} V$. The
equivalence between Eqs. (\ref{me2}) and (\ref{iime}) follows
because $\tilde{\bm{S}}^2 = S_0 \left ( S_0 + 2 \right )$ and
\begin{equation}
\sum_{j=x,y,z} \tilde{S}_j \rho \tilde{S}_j =
\sum_{j,k,\ell=x,y,z} R_{jk} R_{j\ell} S_k \rho
S_\ell = \sum_{j=x,y,z} S_j \rho S_j ,
\end{equation}
where we have used that $R$ is a  rotation and then
\begin{equation}
\sum_{j=x,y,z} R_{jk} R_{j\ell} =
\sum_{j=x,y,z} R^{\rm t}_{kj} R_{j\ell} = (R^{\rm t} R)_{k\ell} =
\delta_{k \ell}.
\end{equation}

\subsection{Steady state}

The steady states of the master equation (\ref{me2}) are given by the solutions to the equation
\begin{equation}
\frac{d \rho_{\rm ss}}{dt}  = 0 \longrightarrow \sum_{j=x,y,z} [
S_j ,[S_j , \rho_{\rm ss} ]] =0 ,
\end{equation}
which is clearly satisfied by the identities $\mathds{1}_n$. Furthermore, if we pretend that \eqref{me2} describes correctly depolarization, we have to show that for any initial state, the asymptotic state of the evolution is a completely unpolarized state $Q(\Omega)=1/(4\pi)$. To that aim, let us first focus on a given subspace $\mathcal{H}_n$. In such a case, first notice that the operators $S_j$ are self-adjoint (and bounded), and there is not a smaller subspace of $\mathcal{H}_n$ invariant under the action of three operators $S_j$, so the representation of $S_j$ on $\mathcal{H}_n$ is irreducible. Therefore, by the Schur's lemma, the only operators commuting with all $S_j$ have to be multiples of the identity. This fulfils the requirements of theorem (5.3) in Ref. \cite{SP80} (see also Ref. \cite{Libro}) and thus in any subspace $\mathcal{H}_n$ the state $\mathds{1}_n/(n+1)$ is the only steady state and any initial state approaches it as $t \rightarrow\infty$.

From this reasoning, the asymptotic state for an arbitrary initial state has to be the form of Eq. \eqref{ss} plus some possible term crossing different subspaces $\mathcal{H}_n$ which does not affect the polarization distribution, so $Q_{\rm ss}(\Omega)=1/(4\pi)$.

\subsection{Evolution of Stokes parameters}

With the above formulas we can obtain once for all the
evolution of the Stokes parameters for every state. This
is because from Eq. (\ref{eeA}) we readily get
\begin{equation}
\frac{d s_0}{dt} =0, \quad
\frac{d s_k}{dt} = - 8 \nu s_k ,
\end{equation}
so that
\begin{equation}
\label{geSp}
s_0 (t) = s_0 (0), \quad
s_k (t) = {\rm e}^{- 8 \nu t} s_k (0),
\end{equation}
and
\begin{equation}
\label{Pst}
P_s (t) = {\rm e}^{- 8 \nu t} P_s (0).
\end{equation}
This universal evolution holds as a consequence of the SU(2)
invariance and agrees with the Mueller matrix for classical
depolarizing systems in Ref. \cite{Pol2}. Actually, the fact that the dynamics of $P_s(t)$ is universal and only depends on its initial value and not on the form of the initial state, makes $P_s(t)$ to be not informative about what states are more robust under depolarization, setting all of them on equal footing. In particular
the Stokes parameters do not distinguish between quantum and
classical states. To find quantum-classical differences we shall
consider the evolution of $P_Q$. This will be done in Sec. V.

\subsection{Evolution of variances of Stokes operators}

A very frequently used measure of uncertainty is variance,
which serves to define basic quantum properties such as
polarization squeezing which is of relevance for metrological
applications \cite{LK,BM,noon}. Using Eq. (\ref{eeA}) we derive the evolution equation for
the mean value of the square of any Stokes-operator component
$S_m = \bm{m} \cdot \bm{S}$, where $\bm{m}$ is any unit
real vector $\bm{m}^2 = 1$,
\begin{equation}
\frac{d \langle S_m^2 \rangle}{dt} = - 24 \nu \langle S_m^2
\rangle + 8 \nu \langle S_0 \left ( S_0 + 2 \right ) \rangle,
\end{equation}
which can be integrated to give
\begin{equation}
\langle S_m^2 \rangle (t) = {\rm e}^{- 24 \nu t} \langle S_m^2 \rangle
(0) + \frac{1}{3} \left ( 1 - {\rm e}^{- 24 \nu t} \right ) \langle
S_0 \left ( S_0 + 2 \right ) \rangle .
\end{equation}
Then, taking into account Eq. (\ref{geSp}) we obtain
\begin{align}
\label{Dsu}
\left ( \Delta S_m \right )^2 (t) &= {\rm e}^{- 24 \nu t} \left (
\Delta S_m \right )^2 (0) \nonumber\\
&+ \frac{1}{3} \left ( 1 - {\rm e}^{- 24
\nu t} \right ) \langle S_0 \left ( S_0 + 2 \right ) \rangle \\
& - \left ( {\rm e}^{- 16 \nu t} - {\rm e}^{- 24 \nu t} \right ) \langle
S_m \rangle^2 (0).\nonumber
\end{align}
We can appreciate that when $t \rightarrow \infty$ we get
$\Delta S_m \rightarrow \langle S_0 \left ( S_0 + 2 \right )
\rangle /3$ for any component $S_m$. A result that agrees well with the imposed
SU(2) invariance.

Alternatively we can express this result in terms of the
symmetric second-order covariance matrix $M = \mathcal{M}-
\mathcal{N}$ with matrix elements
\begin{equation}
\label{socm}
\mathcal{M}_{ij} := \frac{1}{2} \langle (S_i S_j + S_j S_i )
\rangle, \quad \mathcal{N}_{ij} := \langle S_i \rangle
\langle S_j \rangle ,
\end{equation}
such that \cite{RiL}
\begin{equation}
(\Delta S_m )^2 (t) = \bm{m}^{\rm t} M(t) \bm{m}.
\end{equation}
We get  $M (t) = \mathcal{M} (t) -\mathcal{N} (t)$ with
\begin{align}
\label{Mt}
& \mathcal{M} (t) = {\rm e}^{- 24 \nu t} \mathcal{M} (0) + \frac{1}{3}
\left ( 1 - {\rm e}^{- 24 \nu t} \right ) \langle
S_0 \left ( S_0 + 2 \right ) \rangle \mathds{1} , & \nonumber \\
& \mathcal{N} (t) = {\rm e}^{- 16 \nu t} \mathcal{N} (0). &
\end{align}

Next we show that the sum of the uncertainties $(\Delta
\bm{S} )^2$ in Eq. (\ref{svSo}) is a nondecreasing function
of time. From Eqs. \eqref{svSo2} and \eqref{Pst} we get
\begin{equation}
\left ( \Delta \bm{S} \right )^2 (t) = \langle S_0(S_0+2)\rangle - {\rm e}^{- 16 \nu t} P_s^2 (0)\langle S_0\rangle^2 ,
\end{equation}
so that
\begin{equation}
\left ( \Delta \bm{S} \right )^2 (t) > \left ( \Delta \bm{S}
\right )^2 (0) \quad \textrm{for} \;\;\;\langle \bm{S} \rangle
\neq \bm{0} ,
\end{equation}
while
\begin{equation}
\label{cur}
\left ( \Delta \bm{S} \right )^2 (t) = \textrm{constant} ,
\quad \textrm{for} \;\;\;  \langle \bm{S} \rangle = \bm{0} .
\end{equation}

This nondecreasing behavior agrees with common intuition
regarding the effect of random transformations. When approaching
the steady state the sum of the uncertainties reaches its
maximum value $( \Delta \bm{S} )^2 = \langle S_0 \left ( S_0 +
2 \right ) \rangle$ in Eq. (\ref{aab}), i.e. maximum second-order
polarization fluctuations. We stress the curious fact
in Eq. (\ref{cur}) that for vanishing Stokes parameters the
uncertainty does not depend on time. This is because in such a
case $P_s$ is time independent since from Eq. (\ref{geSp}) $\langle
\bm{S} \rangle (0) = \bm{0} \rightarrow \langle \bm{S} \rangle (t)
= \bm{0}$.

Nevertheless, note that the uncertainty of a single Stokes
component may be a decreasing function of time. For example, for
$\langle S_m \rangle = 0$ we have that $\Delta S_m$ decreases if
$(\Delta S_m )^2 (0) > \langle S_0 \left ( S_0 + 2 \right )
\rangle/3$ since in such a case the initial uncertainty is larger
than the final uncertainty $(\Delta S_m )^2 (t \rightarrow \infty )
= \langle S_0 \left ( S_0 + 2 \right ) \rangle/3$. This is the case,
for example, of the twin photon-number state $|n,n  \rangle$ since
at $t=0$,
\begin{align}
(\Delta S_x )^2 = (\Delta S_y )^2 &= 2 n (n+1)  \\
&>\frac{1}{3} \langle S_0 \left ( S_0 + 2 \right ) \rangle =
\frac{4}{3}n (n+1).\nonumber
\end{align}

\subsection{Evolution of principal components}

In previous works we have developed the characterization
of angular-momentum fluctuations via the eigenvectors
$\bm{p}$ and eigenvalues $\lambda$ of the symmetric
covariance matrix \cite{RiL} $M_{ij}$, with $M \bm{p}
=  \lambda \bm{p}$. The eigenvalues of $M$ are the principal
variances $(\Delta S_p )^2$ and the eigenvectors are the
principal components $S_p = \bm{p} \cdot \bm{S}$. Next we
study whether these components are invariant under SU(2)
depolarization. To this end let us distinguish between two
cases, $\langle \bm{S} \rangle = \bm{0}$ and $\langle \bm{S}
\rangle \neq \bm{0}$. From Eq. (\ref{geSp}) this classification
is time invariant since $\langle \bm{S} \rangle (0) = \bm{0}
\rightarrow \langle \bm{S} \rangle (t) = \bm{0}$.

For $\langle \bm{S} \rangle = \bm{0}$ we have $M = \mathcal{M}$,
so that after Eq. (\ref{Mt}) the eigenvectors of $M(t)$ and
$M(0)$ are the same, and the principal components are invariant.

For $\langle \bm{S} \rangle \neq \bm{0}$ the principal components may
vary with dynamics. However, we may decompose
$\bm{S}$ into a longitudinal component $S_\| := \bm{S} \cdot
\langle \bm{S} \rangle / |\langle \bm{S} \rangle|$, with
$\langle S_\| \rangle = |\langle \bm{S} \rangle|$, and two
orthogonal transversal components $S_{\perp,k}$, $k=1,2$,
with $\langle S_{\perp, k} \rangle = 0$. The component
$S_\|$ is invariant since after Eq. (\ref{geSp}) we have
that $\langle S_{\perp, k} \rangle (0) = 0$ implies
$\langle S_{\perp, k} \rangle (t) = 0$ for all $t$.
Thus, if $S_\|$ is a principal component at
$t=0$ we get that the principal components are invariant
since in the transversal subspace spanned by $S_{\perp,k}$
we can apply the same reasoning as the case $\langle \bm{S}
\rangle = \bm{0}$ to the restriction of $M$ to such subspace.

\subsection{Evolution equation for the polarization
distribution}

From Eq. (\ref{me}) it is possible to derive an evolution equation for
the polarization distribution (\ref{SUQ}).
It is convenient to use a slightly different parametrization in the form
\begin{equation}
Q (\bm{\Omega}) =  \sum_{n=0}^\infty \frac{n+1}{4 \pi}
\frac{q(n,\xi)}{( 1 + |\xi |^2)^n},
\end{equation}
where $q(n,\xi) = \langle n,\xi | \rho | n, \xi \rangle$,
and $| n, \xi \rangle$ are the unnormalized SU(2) coherent
states
\begin{equation}
| n, \xi \rangle := \sum_{m=0}^n \begin{pmatrix} n \\ m \end{pmatrix}^{1/2}
\xi^m |m , n-m \rangle ,
\quad
\xi := \cot \frac{\theta}{2} {\rm e}^{-{\rm i} \phi} .
\end{equation}
Next we transform the master equation (\ref{me}) into a
partial differential equation for $q(n,\xi)$. Taking into account that
\cite{BM}
\begin{equation}
\label{spm}
S_+ |n,\xi \rangle = 2 \frac{\partial}{\partial \xi} |n,\xi
\rangle , \quad
S_- |n,\xi \rangle = 2 \xi \left ( n - \xi
\frac{\partial}{\partial \xi} \right ) |n,\xi \rangle ,
\end{equation}
where
\begin{equation}
S_{\pm} = S_x \pm {\rm i} S_y, \quad S_+ = 2 a^\dagger_1 a_2,
\quad S_- = 2 a^\dagger_2 a_1 ,
\end{equation}
and
\begin{equation}
\label{sz}
S_z |n,\xi \rangle = \left ( 2 \xi \frac{\partial}{\partial
\xi} - n \right ) |n,\xi \rangle,
\end{equation}
we obtain
\begin{align}
\label{efp}
\frac{d}{dt} q (n,\xi) &= 4 \nu \left [ n \left ( n |\xi|^2 -1
\right ) + \left ( 1 + |\xi|^2 \right )^2 \frac{\partial}{\partial
\xi} \frac{\partial}{\partial \xi^\ast} \right . & \nonumber \\
& \left .  - n \left ( 1 + |\xi|^2 \right ) \left ( \xi
\frac{\partial}{\partial \xi} + \xi^\ast \frac{\partial}{\partial
\xi^\ast} \right ) \right ] q(n,\xi). &
\end{align}
Despite this equation may be useful in some cases, we do not employ it in the forthcoming sections.

\subsection{Another master equation}

To conclude this section, it is worth comparing the above approach leading to Eqs. (\ref{me}), and (\ref{me2}) with a similar master equation previously considered in Ref. \cite{BAB}. There, the master equation is obtained modeling the depolarization process via light interacting with an atomic reservoir which irreversibly decays because of an additional electromagnetic environment. Under several assumptions the evolution of the light state is given by a master equation of the form
\begin{align}
\label{meb}
 \frac{d \rho}{dt}  &= \gamma_0 \mathcal{L} (S_0) \rho +
\gamma \mathcal{L} (S_+) \rho + \gamma \mathcal{L} (S_-)
\rho  & \nonumber \\
& = \gamma_0 \mathcal{L} (S_0) \rho +
2 \gamma \mathcal{L} (S_x) \rho + 2 \gamma \mathcal{L} (S_y)
\rho , &
\end{align}
where $\gamma_0$ and $\gamma$ are positive constants and
\begin{equation}
\mathcal{L} (A) \rho := 2 A \rho A^\dagger - A^\dagger A \rho
- \rho  A^\dagger A .
\end{equation}

The first factor $ \mathcal{L} (S_0)$ is not relevant for polarization. Therefore, the main difference between both approaches arises from the factor depending on $S_z$ in Eq. (\ref{me}) which is missing in Eq. (\ref{meb}). This implies that the master equation Eq. (\ref{meb}) does not describe SU(2)-invariant depolarization but may well represent some other physical process with different symmetries.

\section{Depolarization of some relevant light states}

Let us particularize the above approach to some simple
but relevant examples. We consider first the paradigmatic case of single-photon states.

\subsection{One-photon states}

The one-photon subspace $\mathcal{H}_1$ is spanned by the
photon-number states
\begin{equation}
 | 1, 0 \rangle = \begin{pmatrix}1 \\ 0 \end{pmatrix},
\quad
 | 0, 1 \rangle = \begin{pmatrix}0 \\ 1 \end{pmatrix},
\end{equation}
being equivalent to an angular momentum $j=1/2$. In the above
basis the most general state can be expressed as
\begin{equation}
\label{1pr}
\rho = \frac{1}{2} \mathds{1}_1 + \frac{1}{2} \bm{s} \cdot \bm{\sigma} ,
\end{equation}
where $\mathds{1}_1$ is the $2 \times 2$ identity matrix, $\bm{\sigma}$
are the three Pauli matrices, and $\bm{s} = \textrm{Tr} \left (
\rho \bm{S} \right )$ are the Stokes parameters being $s_0 = 1$.
In this case $\bm{s}$ and $Q (\bm{\Omega})$ provide the same
information since $\bm{s}$ determine completely both $\rho$ and
$Q(\bm{\Omega})$,
\begin{equation}
Q(\bm{\Omega}) = \frac{1}{4 \pi} \left ( 1 + \bm{s} \cdot
\bm{\bm{\Omega}} \right ).
\end{equation}

In particular, for the degree of polarization $P_Q$ in Eq. (\ref{nddp})
we get
\begin{equation}
D = \frac{1}{3} \bm{s}^2 ,
\quad
P_Q = \frac{\bm{s}^2}{3 + \bm{s}^2} =
\frac{P_s^2}{3 + P_s^2} .
\end{equation}

Taking into account Eq. (\ref{geSp}) we obtain a very simple evolution
for $\bm{s}$
\begin{equation}
\frac{d \bm{s}}{d t} = - 8\nu\bm{s},
\quad
\bm{s} (t) = {\rm e}^{-8\nu t} \bm{s} (0),
\end{equation}
so that
\begin{equation}
\label{1prt}
\rho (t) = \frac{1}{2} \mathds{1}_1 + \frac{1}{2} {\rm e}^{ - 8\nu t} \bm{s} (0)
\cdot \bm{\sigma} ,
\end{equation}
and
\begin{equation}
P_s (t) = {\rm e}^{-8\nu t} P_s (0) , \quad
P_Q (t) = \frac{P_s^2 (0) {\rm e}^{-16\nu t}}{3 + P_s^2 (0)
{\rm e}^{- 16\nu t}} .
\end{equation}

Equation (\ref{1prt}) allows to write this dynamics in the common form of a depolarization channel \cite{Nielsen,dch} widely used in quantum information contexts,
\begin{equation}
\label{1pdch}
\rho (t) = [1-p(t)] \rho (0) + p(t) \rho_{\mathrm{unpol}},
\end{equation}
where
\begin{equation}
\rho_{\mathrm{unpol}} = \frac{1}{2} \mathds{1}_1 , \quad
p(t) = 1- {\rm e}^{-8\nu t} .
\end{equation}

\subsection{Two-photon states}

Next we focus on two-photon systems. This is interesting since it is the simplest subspace including classical and nonclassical polarization states. Thus, we will examine whether typical nonclassical states depolarize faster than classical ones.

The two-photon subspace $\mathcal{H}_2$ is spanned by the photon-number
states
\begin{equation}
 | 2, 0 \rangle = \begin{pmatrix} 1 \\ 0 \\ 0 \end{pmatrix},
\quad
 | 1, 1 \rangle = \begin{pmatrix} 0 \\ 1 \\ 0 \end{pmatrix},
\quad
 | 0, 2 \rangle = \begin{pmatrix} 0 \\ 0 \\ 1 \end{pmatrix},
\end{equation}
being equivalent to an angular momentum $j=1$. In the above
basis the most general state can be expressed as
\begin{equation}
\label{2pr}
\rho = \frac{1}{3} \mathds{1}_3 + \frac{1}{2} \bm{\mu} \cdot
\bm{\Lambda} ,
\end{equation}
where $\mathds{1}_3$ is the $3 \times 3$ identity matrix,
$\bm{\Lambda}$ are the eight Gell-Mann matrices \cite{Pol}
\begin{eqnarray}
\mathbf{\Lambda} _1 = \begin{pmatrix} 0&1&0 \\ 1&0&0 \\ 0&0&0 \end{pmatrix} ,
 & &
\mathbf{\Lambda} _2 = \begin{pmatrix} 0&-{\rm i}&0 \\ {\rm i}&0&0 \\ 0&0&0 \end{pmatrix} ,
\nonumber \\
\mathbf{\Lambda} _3 = \begin{pmatrix} 1&0&0 \\ 0&-1&0 \\ 0&0&0 \end{pmatrix},
& &
\mathbf{\Lambda} _4 = \begin{pmatrix} 0&0&1 \\ 0&0&0 \\ 1&0&0 \end{pmatrix},
\nonumber \\
\mathbf{\Lambda} _5 = \begin{pmatrix} 0&0&-{\rm i} \\ 0&0&0 \\ {\rm i}&0&0 \end{pmatrix},
& &
\mathbf{\Lambda} _6 = \begin{pmatrix} 0&0&0 \\ 0&0&1 \\ 0&1&0 \end{pmatrix},
\nonumber \\
\mathbf{\Lambda} _7 = \begin{pmatrix} 0&0&0 \\ 0&0&-{\rm i} \\ 0& {\rm i}&0 \end{pmatrix},
& &  \mathbf{\Lambda} _8 = \frac{1}{\sqrt{3}}
\begin{pmatrix} 1&0&0 \\ 0&1&0 \\ 0&0&-2 \end{pmatrix} ,
\end{eqnarray}
and $\bm{\mu}$ are eight real parameters. By inserting
Eq. (\ref{2pr}) into Eq. (\ref{me}) and taking into account
the trace orthogonality of Gell-Mann matrices,
\begin{equation}
\textrm{Tr} \left ( \mathbf{\Lambda}_j \mathbf{\Lambda}_k
\right ) = 2 \delta_{jk},
\end{equation}
we obtain an evolution equation for the parameters $\bm{\mu}$:
\begin{equation}
\label{eefm}
\frac{d \bm{\mu}}{d t } = -  \Gamma  \bm{\mu},
\quad
\bm{\mu} (t) = {\rm e}^{- \Gamma t} \bm{\mu} (0),
\end{equation}
where in general for $n$ photons
\begin{equation}
\Gamma_{jk} := - \frac{\nu}{2} \sum_{\ell =x,y,z} \mathrm{Tr} \left (
\Lambda_j S_\ell \Lambda_k S_\ell \right ) + 2 \nu n(n+2) \delta_{jk},
\end{equation}
leading for $n=2$ to
\begin{equation}
\label{M}
\Gamma = 4\nu \begin{pmatrix} 4 & 0 & 0 & 0 & 0 & -2 & 0 & 0 \\
             0 & 4 & 0 & 0 & 0 & 0 & - 2 & 0 \\
             0 & 0 & 5 & 0 & 0 & 0 & 0 & -\sqrt{3}\\
             0 & 0 & 0 & 6 & 0 & 0 & 0 & 0 \\
             0 & 0 & 0 & 0 & 6 & 0 & 0 & 0 \\
            -2 & 0 & 0 & 0 & 0 & 4 & 0 & 0 \\
             0 & -2 & 0 & 0 & 0 & 0 & 4 & 0 \\
             0 & 0 & -\sqrt{3} & 0 & 0 & 0 & 0 & 3 \end{pmatrix}.
\end{equation}
This split into some invariant subspaces
\begin{eqnarray}
& \frac{d}{d t } \begin{pmatrix}\mu_1 \\ \mu_6 \end{pmatrix} = - 8\nu
\begin{pmatrix} 2 & -1 \\ -1 & 2 \end{pmatrix} \begin{pmatrix} \mu_1 \\ \mu_6 \end{pmatrix}, &
\nonumber \\
& \frac{d}{dt} \begin{pmatrix} \mu_2 \\ \mu_7 \end{pmatrix}= - 8\nu
\begin{pmatrix} 2 & -1 \\ -1 & 2  \end{pmatrix} \begin{pmatrix} \mu_2 \\ \mu_7 \end{pmatrix}, &
\end{eqnarray}
\begin{eqnarray}
& \frac{d}{d t} \begin{pmatrix} \mu_3 \\ \mu_8 \end{pmatrix} = - 4\nu
\begin{pmatrix} 5 & -\sqrt{3}\\ -\sqrt{3} & 3 \end{pmatrix}
\begin{pmatrix} \mu_3 \\ \mu_8 \end{pmatrix}, & \nonumber \\
& \frac{d}{d t} \mu_4  = - 24\nu\mu_4, \quad
\frac{d}{d t} \mu_5  = - 24\nu \mu_5, &
\end{eqnarray}
which simplifies the computation of the evolution by
exponentiation of $2 \times 2$ matrices. Moreover it can be
seen that
\begin{equation}
S_x = \sqrt{2} (\mu_1 + \mu_6), \quad
S_y = \sqrt{2} (\mu_2 + \mu_7), \quad
S_z = \mu_3 + \sqrt{3} \mu_8 .
\end{equation}

The evolution of $P_s$ is already determined
from Eq. (\ref{geSp}), implying that all states depolarize
at the same speed. To gain further insight we focus on the
degree of polarization $P_Q$ derived from the $Q$ function.
In this case the SU(2) coherent states in the photon-number
basis are
\begin{equation}
| 2,\bm{\Omega} \rangle = \cos^2 \frac{\theta}{2} {\rm e}^{-2 {\rm i}\phi}| 2, 0
\rangle + \frac{\sin \theta}{\sqrt{2}}
{\rm e}^{-{\rm i} \phi} | 1, 1 \rangle + \sin^2 \frac{\theta}{2}
 | 0, 2 \rangle ,
\end{equation}
so that the $Q$ function of the most general state (\ref{2pr})
is of the form
\begin{equation}
\label{Qf2f}
Q (\bm{\Omega}) = \frac{1}{4 \pi} \left [ 1 + \frac{3}{2}
\bm{\mu} \cdot \bm{\lambda} (\bm{\Omega}) \right ] ,
\end{equation}
where
\begin{equation}
\bm{\lambda} (\bm{\Omega}) := \langle 2, \bm{\Omega} |
\bm{\Lambda} | 2, \bm{\Omega} \rangle .
\end{equation}
To compute $P_Q$ we need the integral $\int d \Omega
Q^2 (\bm{\Omega})$ that after Eq. (\ref{Qf2f}) can be expressed
in terms of the $\bm{\mu}$ parameters as
\begin{equation}
D=4 \pi \int d \Omega Q^2 (\bm{\Omega}) -1= \bm{\mu}^{\rm t}
\Phi \bm{\mu},
\end{equation}
so that
\begin{equation}
P_Q = \frac{\bm{\mu}^{\rm t} \Phi \bm{\mu}}{1 + \bm{\mu}^{\rm t} \Phi
\bm{\mu}},
\end{equation}
where $\Phi$ is the $8 \times 8$ matrix
\begin{equation}
\Phi_{jk} := \frac{9}{16\pi}\int d \Omega \lambda_j (\bm{\Omega}) \lambda_k
(\bm{\Omega}) ,
\end{equation}
leading to
\begin{equation}
\Phi  = \frac{3}{20}
    \begin{pmatrix}3 & 0 & 0 & 0 & 0 & 2 & 0 & 0 \\
             0 & 3 & 0 & 0 & 0 & 0 & 2 & 0 \\
             0 & 0 & 2 & 0 & 0 & 0 & 0 & \sqrt{3}\\
             0 & 0 & 0 & 1 & 0 & 0 & 0 & 0 \\
             0 & 0 & 0 & 0 & 1 & 0 & 0 & 0 \\
             2 & 0 & 0 & 0 & 0 & 3 & 0 & 0 \\
             0 & 2 & 0 & 0 & 0 & 0 & 3 & 0 \\
             0 & 0 & \sqrt{3} & 0 & 0 & 0 & 0 & 4 \end{pmatrix} .
\end{equation}
It can be appreciated that $\Phi$ displays the same structure of
invariant subspaces than the evolution matrix $\Gamma$ in Eq. (\ref{M}).

\subsubsection{SU(2) coherent states}

We have computed the degree of polarization for two particular
initial SU(2) coherent states: one at the north pole $| 2,
\bm{\Omega}_{\rm po} \rangle$ with $\theta=0$, and the other one at
the equator $| 2, \bm{\Omega}_{\rm eq} \rangle$, with $\theta = \pi /2$
and $\phi=0$,
\begin{equation}
\label{2su2}
|2, \bm{\Omega}_{\rm po} \rangle = | 2 , 0 \rangle , \quad
|2, \bm{\Omega}_{\rm eq} \rangle = \frac{1}{2} \left ( | 2 , 0
\rangle + \sqrt{2} | 1 , 1 \rangle + | 0 , 2 \rangle
\right ) ,
\end{equation}
for which, at $t=0$ we have the following vectors
$\bm{\mu}$
\begin{eqnarray}
& \bm{\mu}_{\rm po} = \left ( 0, 0, 1, 0, 0, 0, 0, \frac{1}{\sqrt{3}}
\right ) , & \nonumber \\
& \bm{\mu}_{\rm eq} = \left ( \frac{1}{\sqrt{2}}, 0,
- \frac{1}{4}, \frac{1}{2}, 0, \frac{1}{\sqrt{2}}, 0,
\frac{1}{4 \sqrt{3}} \right ) . &
\end{eqnarray}
The evolution of the degree of polarization $P_Q$ is exactly the
same for both states. This can be simply expressed in terms of the
distance to unpolarized light (\ref{dD}) as
\begin{equation}
\label{Dc2}
D(t) =  \frac{3}{4} {\rm e}^{- 16\nu t} + \frac{1}{20} {\rm e}^{-48\nu t}.
\end{equation}
The evolution of the degree of polarization normalized to its
initial value $P_Q (t) /P_Q (0)$ is represented in Fig. 1 as
a blue line. We stress that because of the explicit SU(2) symmetry of the evolution
and of the definitions of  degree of polarization, all SU(2) coherent states have he same
$D(t)$, since they are connected by SU(2) transformations.
\begin{figure}
\begin{center}
\includegraphics[width=\columnwidth]{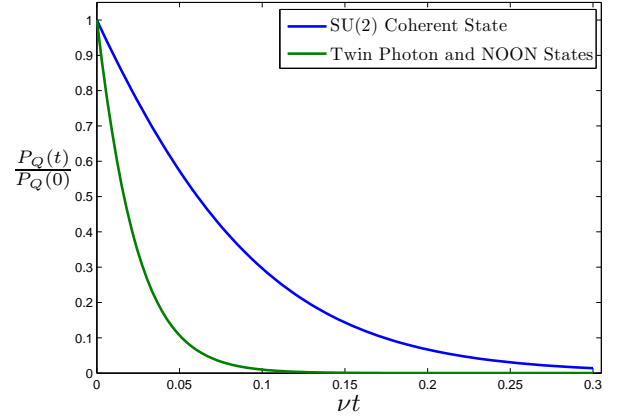}
\end{center}
\caption{Plot of the evolution of the degree of polarization normalized
to its initial value $P_Q (t) /P_Q (0)$ as a function of time computed for two-photon SU(2)
coherent states [blue (upper) line] and twin-photon and NOON states [green
(lower) line].}
\end{figure}

\subsubsection{Twin-photon and NOON States}

Let us compare the above depolarization of SU(2) coherent
states with the effect of depolarization of nonclassical
states, such as, in the number basis
\begin{equation}
\label{2ss}
| \xi \rangle = | 1 , 1 \rangle , \quad
| \zeta \rangle = \frac{1}{\sqrt{2}} \left ( | 2 , 0 \rangle
+ | 0, 2 \rangle \right ) ,
\end{equation}
with
\begin{eqnarray}
& \bm{\mu}_\xi = \left ( 0, 0, -1, 0, 0, 0, 0,
\frac{1}{\sqrt{3}} \right ), & \nonumber \\
& \bm{\mu}_\zeta = \left ( 0, 0, \frac{1}{2}, 1, 0, 0, 0,
- \frac{1}{2 \sqrt{3}} \right ). &
\end{eqnarray}
The vectors $| \xi \rangle $, $| \zeta \rangle $ are
eigenstates of $S_z$, $S_y$, respectively, with 0 eigenvalue. Thus, they are SU(2) equivalent since there is a SU(2) transformation
$U$ such that $| \xi \rangle = U | \zeta \rangle$. This is a
rotation $R$ of Stokes operators of angle $\pi/2$ around axis $x$.
Moreover, both can be regarded as the limit of SU(2) squeezed
states at infinite squeezing, since larger squeezing implies smaller $\bm{s}$ \cite{RiL},
$\bm{s} =0$ for both states. Furthermore  $| \zeta \rangle$ is also a weak version of
the Schr\"{o}dinger cat state \cite{LK,BM,noon}.

The evolution of the degree of polarization can be simply
expressed in terms of the distance to unpolarized light (\ref{dD})
as
\begin{equation}
\label{Ds}
D(t) = \frac{1}{5} {\rm e}^{- 48\nu t } .
\end{equation}
The evolution of the degree of polarization normalized to its
initial value $P_Q (t) /P_Q (0)$ is represented in Fig. 1 as
a green line.

In Fig. 1 and Eqs. (\ref{Dc2}) and (\ref{Ds}) we can clearly
appreciate that these nonclassical states depolarize three times
faster than the SU(2) coherent states that are classical
regarding polarization \cite{aclaracion}. This agrees with the common idea that
nonclassical states are more sensitive to randomness and other
imperfections than classical ones.

\begin{figure}
\begin{center}
\includegraphics[width=\columnwidth]{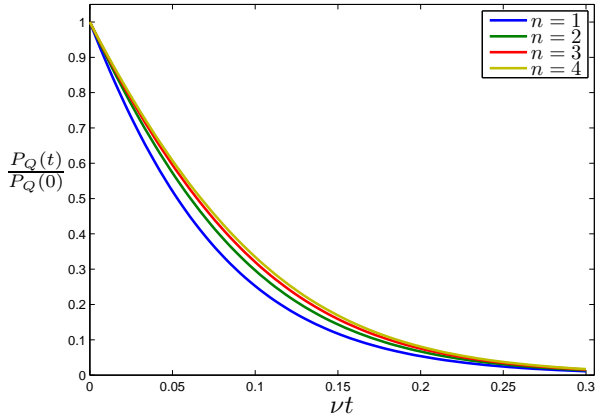}
\end{center}
\caption{Evolution of the degree of polarization normalized to its initial value
$P_Q (t) /P_Q (0)$ as a function of time computed SU(2) coherent states for several photon numbers. The upper line corresponds to $n = 4$ photons, the next line to $n = 3$, and so on.}
\end{figure}

\subsection{Higher photon states}

The evolution of the degree of polarization in subspaces with a higher number of photons can be calculated following a
 similar strategy as for two photon states, since for any $\mathcal{H}_n$, the state of light may be written as
\begin{equation}
\rho=\frac{ \mathds{1}_n}{n+1} + \frac{1}{2}
\bm{\mu} \cdot \bm{\Lambda},
\end{equation}
where generally the matrices $\bm{\Lambda}$ are the generators of the Lie algebra $\mathfrak{su}(n + 1)$.

For the sake of illustration, we have computed the dynamics of the degree of polarization for SU(2) coherent states and the nonclassical states \eqref{2ss} for three and four photons.

\subsubsection{SU(2) coherent states}

For three photons the evolution of the distance to unpolarized light of SU(2) coherent states becomes
\begin{equation}
D(t)= \frac{27}{25} {\rm e}^{-16 \nu t}+ \frac{1}{5}{\rm e}^{-48 \nu t}+ \frac{1}{175}{\rm e}^{-96 \nu t},
\end{equation}
and for four photons
\begin{equation}
D(t)=\frac{4}{3} {\rm e}^{-16 \nu t}+ \frac{20}{49} {\rm e}^{-48 \nu t}+\frac{1}{28} {\rm e}^{-96 \nu t}+\frac{4}{3087} {\rm e}^{-160 \nu t}.
\end{equation}
Hence, the rate of depolarization is approximately independent of the number of photons. This can be seen in Fig. 2, where we have compared
the depolarization of SU(2) coherent states with one, two, three, and four photons. Moreover, this is consistent with the evolution form in Eq. (\ref{sevo}) since, roughly speaking, $D(t)$ is proportional to $\rho^2$ so that the time dependence must be a combination of the decaying  factors $\exp [-8 k(k+1) \nu t]$  for  $k=1, \ldots n$, the $k=0$ term being absent since $D(t)
\rightarrow 0$ as $t \rightarrow \infty$.

\subsubsection{Twin-photon and NOON States}
The rate of depolarization of nonclassical states with the number of photons is also approximately constant. For three photons, there are not twin-photon states, for NOON states we have
\begin{equation}
D(t)= \frac{1}{5}e^{-48 \nu t}+\frac{2}{35}{\rm e}^{-96 \nu t}.
\end{equation}
Similarly, in the case of four photons we obtain the same result for twin-photon and NOON states,
\begin{equation}
D(t)=\frac{20}{49} {\rm e}^{-48 \nu t}+\frac{29}{1372}{\rm e}^{-160 \nu t}.
\end{equation}

\subsection{Depolarization channel}

In general, for more than one photon states, the depolarization dynamics given by Eq. \eqref{me} cannot be written with the simple formula for depolarization channels (\ref{1pdch}), except in some particular cases.
These are the eigenvectors of $\Gamma$, $\Gamma \bm{\mu} = \eta
\bm{\mu}$ (or their superpositions for the same eigenvalue
$\eta$), since, in such a case, we have
\begin{equation}
\rho (t) = \frac{ \mathds{1}_n}{n+1} + \frac{1}{2} {\rm e}^{- \eta t}
\bm{\mu} (0) \cdot \bm{\Lambda} = p(t) \rho_{\mathrm{unpol}}
+ [1-p(t)] \rho (0),
\end{equation}
with
\begin{equation}
\rho_{\mathrm{unpol}} = \frac{1}{n+1} \mathds{1}_n , \quad
 p(t)= 1-\exp(-\eta t).
\end{equation}
For two photon states, it can be seen that this is precisely the case of the states (\ref{2ss}).

\section{Conclusions}

In this work, we have developed a SU(2)-invariant approach to the depolarization of quantum states of light as the effect of random unitary SU(2) transformations. By considering their infinitesimal form under the assumption of Markovianity we have derived an associated SU(2)-invariant master equation and analyzed its main properties.

We have applied this formalism to several quantum states showing that the behavior of the simplest (first moments) degree of polarization is independent of the initial state. However, more complete degrees of polarization allows us to assert that relevant nonclassical states depolarize faster
than classical ones. Moreover, the rate of depolarization is approximately independent of the number of photons.

Finally, we have analyzed the compatibility of our approach to depolarization with the usual form of depolarization channels. We have pointed that both proposals are equivalent for one photon states but not in general. Hence, the model considered here provides a non-trivial generalization of depolarization channels to arbitrary photon numbers.

\section*{Acknowledgements}

We acknowledge financial support from Spanish MINECO
grants FIS2009-10061, FIS2012-33152, FIS2012-35583, CAM research consortium QUITEMAD S2009-ESP-
1594 and UCM-BS grant GICC-910758.


\begin{thebibliography}{00}
\bibitem{UN}
A map $\mathcal{E}$ is said to be unital if $\mathcal{E}(\mathds{1})=\mathds{1}$.
\bibitem{aclaracion} For SU(2) squeezed coherent states with a finite amount of squeezing the decay of the polarization lies between the blue and green lines in Fig. 1. Since for infinite squeezing $\bm{s}$ vanishes, so does the component ${\rm e}^{-16\nu t}$ in $D(t)$ because of the arguments given in Sec. IV. A.
\end{thebibliography}
\end{document}